\documentclass[]{jfm}

\usepackage{graphicx}
\usepackage{newtxtext}
\usepackage{newtxmath}
\usepackage{natbib}
\usepackage{hyperref}
\usepackage{color}
\usepackage[normalem]{ulem}
\DeclareRobustCommand{\erase}{\bgroup\markoverwith{\textcolor{red}{\rule[.5ex]{2pt}{1pt}}}\ULon}
\DeclareMathAlphabet{\mathcal}{OMS}{cmsy}{m}{n}

\hypersetup{
    colorlinks = true,
    urlcolor   = blue,
    citecolor  = black,
}

\newcommand{\RomanNumeralCaps}[1]

\title{G\"ortler number-based scaling of boundary-layer transition on rotating cones in axial inflow}

\author{Sumit  Tambe\aff{1}\aff{2}
  \corresp{\email{sumittambe@iisc.ac.in}},
  Kentaro Kato\aff{3}
 \and Zahir Hussain\aff{4}} 

\affiliation{\aff{1}Aerospace Engineering, Indian Institute of Science, Bangalore, 560012, India.
\aff{2}Aerospace Engineering, Delft University of Technology, 2629HS Delft, The Netherlands.
\aff{3} Department of Mechanical Systems Engineering, Shinshu University, Nagano, Japan. 
\aff{4} Aerospace Engineering,
School of Engineering, University of Leicester, University Road, Leicester LE1 7RH, United Kingdom.}

\begin{document}
\maketitle

    \begin{abstract}{This paper reports on the efficacy of the G\"ortler number in scaling the laminar-turbulent boundary-layer transition on rotating cones facing axial inflow. Depending on the half-cone angle $\psi$ and axial flow strength, the competing centrifugal and crossflow instabilities dominate the transition. Traditionally, the flow is evaluated by using two parameters: the local meridional Reynolds number $Re_l$ comparing the inertial versus viscous effects and the local rotational speed ratio $S$ accounting for the boundary-layer skew. We focus on the centrifugal effects, and evaluate the flow fields and reported transition points using G\"ortler number based on the azimuthal momentum thickness of the similarity solution and local cone radius. The results show that G\"ortler number alone dominates the late stages of transition (maximum amplification and turbulence onset phases) for a wide range of investigated $S$ and  half-cone angle ($15^{\circ} \leq \psi \leq 50^{\circ}$), although the early stage (critical phase) seems to be not determined by the G\"ortler number alone on the broader cones ($\psi=30^{\circ}$ and $50^{\circ}$) where the primary crossflow instability dominates the flow. Overall, this indicates that the centrifugal effects play an important role in the boundary-layer transition on rotating cones in axial inflow.}
   
\end{abstract}


{\bf MSC Codes }  {\it(Optional)} Please enter your MSC Codes here
\section{Introduction}
\label{sec:introduction}

Rotating cones in axial inflow are one of the simplified models for probing the transition phenomena in three-dimensional boundary layers developed on aero-engine-nose-cones, launch vehicle tips, turbo-machinery rotors, etc. Generally, the rotation destabilises the boundary layer on a rotating cone such that the disturbances grow to form coherent vortex structures close to the cone surface. In the presence of the meridional velocity component, the instability-induced vortices align in a spiral vortex pattern around the cone surface \citep{Kobayashi1983a}. As the azimuthal wall velocity increases with the radius, the spiral vortices grow and eventually set on the turbulence \citep{Kohama1984a}. In practice, this transition phenomena, i.e. the spiral vortex growth and the turbulence onset, affect the performance of an engineering system---by altering the momentum distribution near the wall, affecting the local skin friction and heat transfer. 

The cone rotation destabilises the boundary layer through two types of primary instabilities: centrifugal and crossflow instabilities. The centrifugal instability relates to the balance between the centripetal force and the radial pressure gradient---inducing counter-rotating vortices on rotating cylinders \citep{Taylor1923, Serre2023}, concave walls \citep{Gortler1954} and rotating cones with relatively small half-apex angle $\psi \lesssim 30^\circ$ \citep{Kobayashi1983} in still fluid \citep{Hussain2012, Hussain2014} and in axial flow \citep{Hussain2016,Song_Dong_2023,Song_Dong_Zhao_2023}. The crossflow instability arises from the inflectional meridional velocity profile---inducing co-rotating vortices, which have been investigated on rotating disks \citep{Smith1947,Gregory1955, Lingwood1995}, on smooth rotating broad cones $\psi \gtrsim 30^\circ$ \citep{Kobayashi1983a} within still fluid \citep{Garrett2009}, axial flow \citep{Garrett2010}, and, more recently, on rough rotating cones \citep{almalki2022}, as well as on swept wings \citep{Kohama2000}. 

Generally, a rotating-cone boundary layer undergoes transitions through three distinctly observable phases along the cone: 1) the \textit{critical} phase where the instability-induced spiral vortices begin their growth, 2) the \textit{maximum amplification} phase where the spiral vortex amplification peaks, where the vortices rapidly enhance mixing of the outer and inner flow, and 3) the \textit{turbulence onset} phase where the velocity fluctuation spectra start resembling a general turbulence spectrum \citep{Kobayashi1983,Kohama1984}. The present article refers to these phases as transition points in geometric or flow-parameter spaces. In the transition region from the critical to turbulence onset, the instability-induced spiral vortices alter the thermal footprint and velocity distribution of the cone boundary layer. These effects are measurable and useful in identifying the phases of boundary-layer transition on rotating cones, as reported in the previous literature \citep{Kobayashi1983,Kobayashi1987,Tambe2021,Kato2021}. 

On a cone/disk rotating in still fluid, transition has been evaluated by a single parameter such as rotational Reynolds number based on the local radius and wall velocity \citep{Lingwood1995,Lingwood1996,Kobayashi1983a, Garrett2009, Hussain2014}. Recently, \citet{Kato2021} suggested another parameter, G\"ortler number $G$ and experimentally showed that thickening of the boundary layer due to transition can be scaled by G\"ortler number rather than the Reynolds number on a $\psi=30^\circ$ cone.

\begin{figure}
    \centering
    \includegraphics[width=0.65\textwidth]{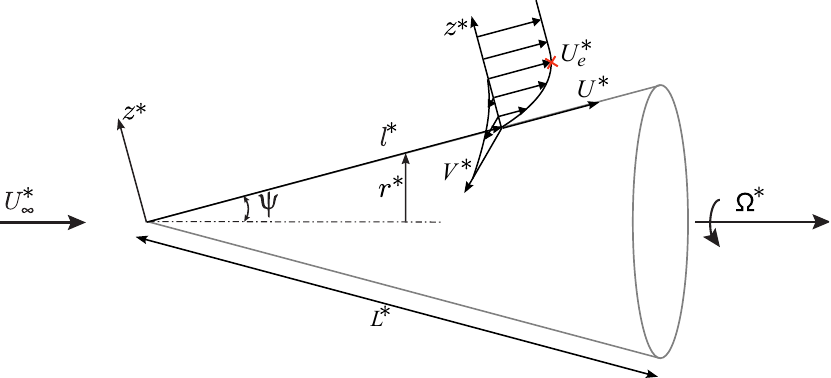}
    \caption{Schematic of a rotating cone in axial inflow.}
    \label{fig:Cone Sketch}
\end{figure}

When the cone is rotating in axial inflow, however, both cone rotation and axial inflow are two independent control parameters. Therefore, \cite{Kobayashi1983,Kobayashi1987} identified two flow parameters for scaling the boundary-layer transition on rotating cones in axial inflow: 1) the local rotational speed ratio $S=r^*\Omega^*/U_e^*$ which accounts for the boundary-layer skew, and 2) the local Reynolds number $Re_l=l^*U_e^*/\nu^*$ comparing the inertial versus viscous effects. Here, asterisk denotes dimensional variables; as schematically shown in figure \ref{fig:Cone Sketch}, $r^*=l^*\sin\psi$ is the local cone radius, $l^*$ is the local meridional length from the cone apex, $\Omega^*$ is the angular velocity of the cone, $U_e^*$ is the meridional component of the boundary layer edge velocity, $\nu^*$ is the kinematic viscosity, $U^*_{\infty}$ is the free stream velocity, $L^*$ is the total meridional length of a cone, and $U^*$, $V^*$ are the meridional and azimuthal velocity components.

The present article reports the efficacy of G\"ortler-number-based scaling of the boundary layer transition on rotating cones in axial inflow, which reduces the two-parameter scaling ($Re_l-S$) down to a single parameter $G$. $\S2$ describes the G\"ortler number and the basic flow formulation as functions of the local rotational speed ratio $S$, which inversely represents the axial inflow strength. $\S3$ describes the experiments, in which the flow data were obtained. $\S4$ presents the measured flow data and reported transition points along conventional $Re_l-S$ scale as well as along G\"ortler number scale. $\S5$ concludes the article.

\section{G\"ortler number and the basic flow formulation}
\label{sec:formulation}

\citet{Gortler1954} showed that due to the centrifugal effects, a boundary layer on a concave wall becomes unstable and its instability induces counter-rotating vortices---also known as G\"ortler vortices. Furthermore, the vortices form at a constant G\"ortler number---which is a product of two non-dimensional parameters: Reynolds number comparing the inertial versus viscous effects and a curvature term $\epsilon_c$ accounting for the wall-normal extent of the viscous effects, e.g. boundary-layer thickness compared to the radius of wall curvature \citep{Taylor1923,Gortler1954,Saric1994}. 

On a rotating cone, the centrifugal effects exist because the viscous flow on the cone wall follows the curved motion of the rotating cone surface. The curved flow within the boundary layer affects its instability behaviour which can be evaluated by G\"ortler number $G$ formulated with the azimuthal-momentum-based Reynolds number $Re_{\delta^*}=r^*\Omega^*\delta^*/\nu^*$  and the curvature term $\epsilon_c=\sqrt{\delta^*/r^*}$:

\begin{equation}
    G=Re_{\delta^*}\epsilon_{c}=\frac{r^*\Omega^*\delta^*}{\nu^*}\sqrt{\frac{\delta^*}{r^*}}=\sqrt{\delta^3r}=\sqrt{\delta^3l \sin\psi}. 
    \label{eq:G}
\end{equation}

Here, $\Omega^*$ is the angular velocity and the azimuthal momentum thickness 

\begin{equation}
  \delta^*=\int_0^\infty V(1-V)dz^*. \;\;\;\; 
\end{equation}
Here, V is the azimuthal velocity normalised by the local wall velocity $r^*\Omega^*$, $z^{*}$ is the wall-normal coordinate, $\delta=\delta^*/\delta_\nu^*$ is the azimuthal momentum thickness normalised by the length scale  $\delta_\nu^*=\sqrt{\nu^*/\Omega^*}$, $r=r^*/\delta_\nu^*$, and $l=l^*/\delta_\nu^*$. Since the strongest curvature appears at the rotating cone surface, the local cone radius  $r^*$ is chosen as the radius of curvature in $\epsilon_c$; and as the rotation adds momentum in the azimuthal direction, the azimuthal momentum thickness $\delta^*$ accounts for the wall-normal extent of viscous effects in $\epsilon_c$.

For a fixed non-dimensional radius $r$, G\"ortler number only depends on the non-dimensional azimuthal momentum thickness $\delta$ in equation (\ref{eq:G}). In the present work, we used the basic flow to compute $\delta$ , assuming that the instability-induced distortion of the mean flow remains small until the maximum amplification phase, which will be validated through figures \ref{fig:15meanU} and \ref{fig:30_50meanU} in $\S$4. In the rest of this section, we describe the basic flow formulation and the effect of axial flow (or $S$) on $\delta$, which consequently affects the local G\"ortler number $G$ for a given $r$. 

The basic flow is computed using the formulation given in section 2.3 of \cite{Hussain2010}, which is based on the formulation by \citet{koh1967}. Following the Mangler transformation, Hussain used a stream function-based similarity type transformation to obtain the governing equations in the non-dimensional form \citep{Hussain2010}. The stream function

\begin{equation}
 \Psi= \left(\frac{m+3}{2s^{\frac{1}{2}}} \sin \psi \right)^{-\frac{1}{2}} f(s,\eta_1) \implies U^*=U_e^* \frac{\partial f(s,\eta_1)}{\partial \eta_1} \textrm{, and } V^*=r^*\Omega^*(g(s,\eta_1)+1). 
\end{equation}

 Here, 

\begin{equation}
s= S^2=\left( \frac{r^*\Omega^*}{U_e^*} \right) ^2 \textrm{, } \eta_1=\eta \left( \frac{m+3}{2s^{\frac{1}{2}}}\sin \psi \right)^{\frac{1}{2}},
\label{eq:base flow scales}
\end{equation}
$\eta$ is the scaled wall-normal coordinate $z^*/\delta_\nu^*$, $m$ is the exponent in the potential flow solution over a cone $U_e^*=C^*{l^*}^m$, where $C^*$ is a constant; for $\psi=15^\circ$, $30^\circ$, and $50^\circ$, $m=0.0396$, $0.117$, and $0.3$, respectively \citep{Hussain2010}. The governing partial differential equations of the basic flow are as follows :

\begin{equation}
     f''' + ff''+\frac{2m}{m+3}(1-f'^2)+\frac{2s}{m+3}[(g+1)^2+2(1-m)\left(f''\frac{\partial f}{\partial s}-f'\frac{\partial f'}{\partial s}\right)]=0,
\end{equation}

\begin{equation}
     g''+fg'-\frac{4}{m+3}f'(g+1)+\frac{4(1-m)s}{m+3}\left(g'\frac{\partial f}{\partial s} -f' \frac{\partial g}{\partial s} \right)=0.
\end{equation}

 Here, $'$ denotes the $\partial/\partial \eta_1$. The boundary conditions are:

\begin{equation}
     f=0,~f'=0,~g=0,~\textrm{on } \eta_1=0;
   ~\textrm{and } f' \rightarrow1,~g\rightarrow-1,~ \textrm{as } \eta_1 \rightarrow \infty.
\end{equation}
A commercial routine NAG D03PEF is used to obtain the basic flow solution. \cite{Hussain2016} have used this basic flow to successfully predict the trend of the critical Reynolds number for the instability onset on a rotating slender cone ($\psi=15^\circ$) in axial inflow---agreeing with the experimental results \citep{Kobayashi1983,Tambe2021}, also shown in figure \ref{fig:Re_S_G}(a).

\begin{figure}
    \centering
    \includegraphics[width=0.94\textwidth]{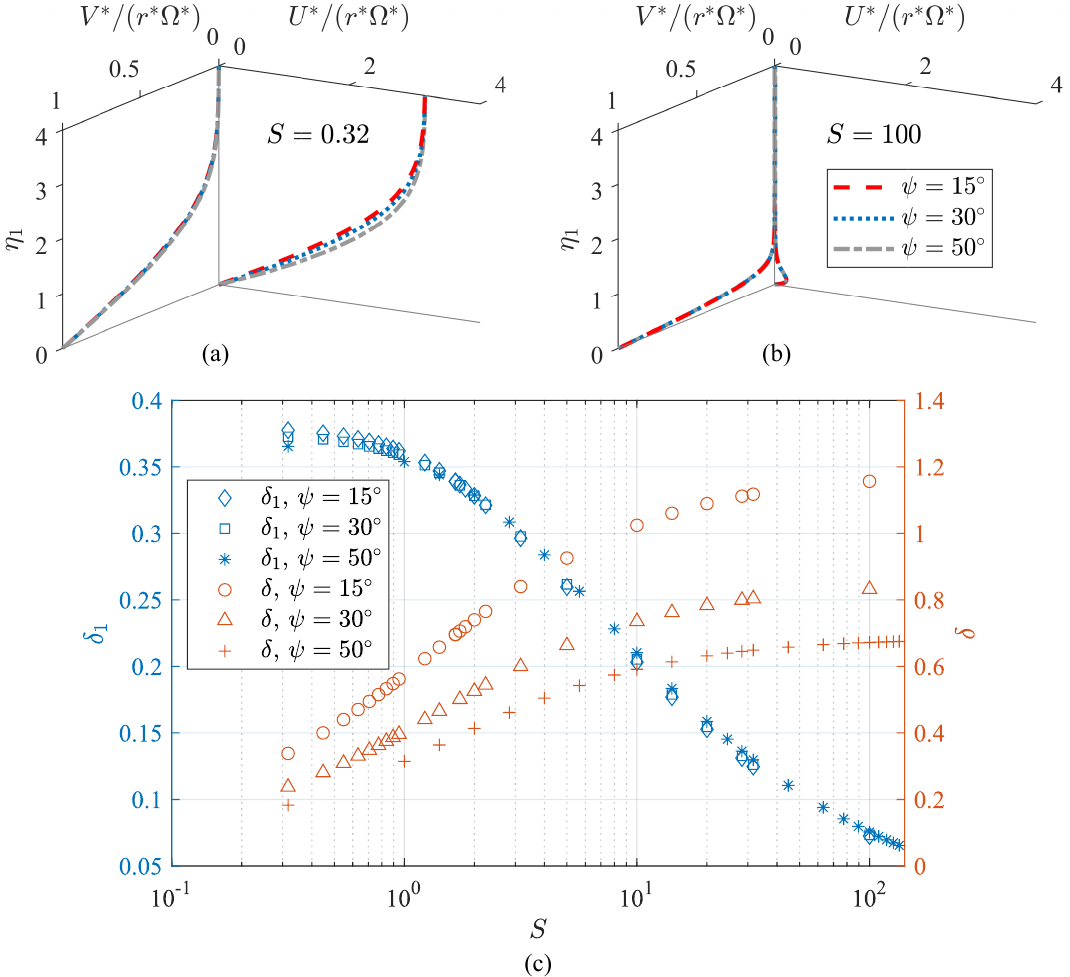}
    \caption{Three-dimensional boundary layer profiles on rotating cones ($\psi=15^\circ$, $30^\circ$, and $50^\circ$) with examples of (a)  strong axial inflow $S=0.32$ and (b) strong rotation $S=100$. (c) Variation of the azimuthal momentum thickness with the local rotational speed ratio $S$. $\delta_1$ is the momentum thickness in the transformed wall-normal coordinate $\eta_1$. $\delta$ is the momentum thickness in the scaled wall normal coordinate $\eta=z^*/\delta_\nu^*$.  }
    \label{fig:basic flow}
\end{figure}

Figure \ref{fig:basic flow} shows the computed velocity profiles of the basic flow for three different half-cone angles $\psi=15^\circ$, $30^\circ$, and $50^\circ$. When the axial inflow is dominant over the rotation (e.g. figure \ref{fig:basic flow}(a) at $S=0.32$), the momentum is distributed in both azimuthal and meridional directions. However, when the rotation is dominant (e.g. figure \ref{fig:basic flow}(b) at $S=100$), most of the momentum is distributed in the azimuthal direction. Furthermore, the meridional velocity takes the inflectional form, owing to the increased meridional pressure gradient caused by the strong rotation.

Owing to the transformation (equation (\ref{eq:base flow scales})), the basic flow profiles for different half-cone angles $\psi$ closely follow each other, especially in the azimuthal direction, see figures \ref{fig:basic flow}(a) and (b). Consequently, their azimuthal momentum thickness $\delta_1$ (in $\eta_1$ coordinates) follows a common trend with respect to the local rotational speed ratio $S$, see figure \ref{fig:basic flow}(c). This shows that varying axial inflow influences the wall-normal distribution of the azimuthal momentum. When transformed back onto the physical coordinates $\eta$, azimuthal momentum thickness $\delta$ increases with $S$ and decreases with the half-cone angle $\psi$. Thus, at a fixed non-dimensional radius $r$, increasing axial inflow strength or reducing $S$ will lower the local G\"ortler number $G$, weakening the centrifugal effects. This shows that the G\"ortler number formulation in equation (\ref{eq:G}) accounts for the axial flow strength (inverse of local rotational speed ratio $S$) through $\delta$.

\section{Methodology}

The efficacy of G\"ortler number-based scaling for boundary layer transition on rotating cones is assessed by using surface temperature fluctuations and the velocity fields obtained as described in \citet{Tambe2022a} and by evaluating the G\"ortler numbers for the transition points reported along the two parameter $Re_l-S$ space in the literature. Parts of the raw data have been used to estimate the transition points reported by \cite{Tambe2021,Tambe2023}.

The experiments were performed at a low-speed open jet wind tunnel (named W-tunnel) at Aerospace Engineering, Delft University of Technology. The cones, made of polyoxymethylene (POM), were rotated in an axial inflow with the free stream velocity $U_{\infty}^*=0.7-10.7$~m/s and typical turbulence level below $0.01U_{\infty}^*$.
Infrared thermography (IRT) is performed at a frequency of  $200$~Hz using an infrared camera \textit{FLIR (CEDIP) SC7300 Titanium} to detect the thermal footprints of the instability-induced features, as described in \cite{Tambe2019a,Tambe2021}. Moreover, the meridional velocity field is measured with two component particle image velocimetry (PIV) at a frequency of $2000$~Hz, using a high-speed camera \textit{Photron Fastcam SA-1} and a high-speed laser \textit{Nd:YAG Quantronix  Darwin Duo 527-80-M}. The flow is seeded with smoke particles having a mean diameter of around $1\mu$m. Two-component velocity vector fields (with the vector pitch of around $2.6 \times 10^{-4} ~ \textrm{m}$) are obtained using a commercial software \textit{DaVis 8.4.0}.

Cones with different half-cone angles $\psi=15^\circ$, $30^\circ$, and $50^\circ$ are rotated at various rotational speeds ($0-13500$ r.p.m.) to obtain different combinations of the operating conditions, i.e. $S_b=L^*\sin \psi \Omega^* /U_{\infty}^*$ and inflow Reynolds number $Re_L=L^*U_{\infty}^*/\nu^*$. As $\delta$ varies with $S$ (figure \ref{fig:basic flow}(c)), the measurement uncertainties of $S$ and $l^*$ ($\pm 0.06S$ and $\pm 0.02l^*$, respectively) cause uncertainty in G\"ortler number (using equation (\ref{eq:G})) around $\pm 0.1G$ . 

\section{Results and discussions}

\begin{figure}
    \centering
    \includegraphics[width=1\textwidth]{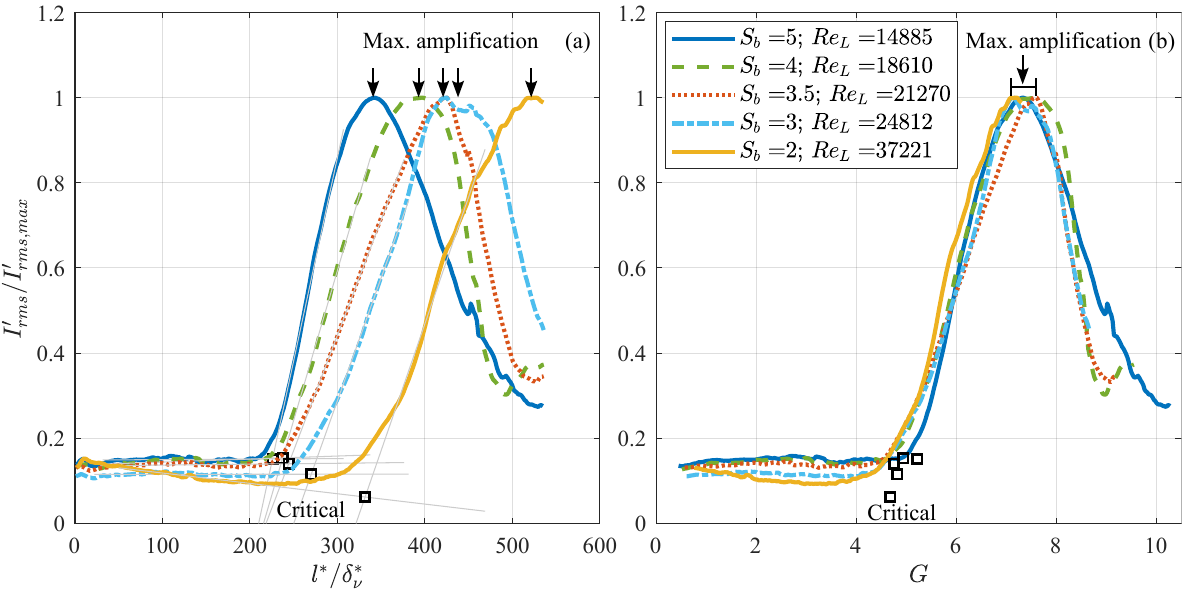}
    \caption{Meridional profiles of r.m.s. of surface temperature fluctuations ($I'_{rms}/I'_{rms,max}$), caused by the growth of instability-induced spiral vortices on a rotating cone ($\psi=15^\circ$), represented along (a) scaled meridional length $l^*/\delta_\nu^*$ and (b) G\"ortler number scale. Squares and downward arrows represent the critical and maximum amplification points, respectively. }
    \label{fig:Irms}
\end{figure}

During the boundary-layer transition on a rotating cone in axial inflow, the growing spiral vortices increase the surface temperature fluctuations, which is detected using infrared thermography (IRT) \citep{Tambe2019a, Tambe2021}. Figure \ref{fig:Irms}(a) shows the r.m.s. of surface temperature fluctuations along a cone with the half-cone angle $\psi=15^\circ$ at different operating conditions, i.e. different combinations of base rotational speed ratio $S_b$ and inflow Reynolds number $Re_L$. The temperature fluctuations are in terms of the normalised pixel intensity $I'_{rms}/I'_{rms, max}$. Critical and maximum amplification phases of the spiral vortex growth are identified for all profiles, examples are marked in figure \ref{fig:Irms}(a). 
Here, the critical points are the intersection points (marked by squares) of the baseline noise level and the least-square linear fit through the rising $I'_{rms}/I'_{rms, max}$, which represents the rapid growth of the spiral vortices. Further downstream, the growth saturates at the $I'_{rms}/I'_{rms, max}$ peak (the maximum amplification phase marked by the arrows), and subsequently, the flow becomes turbulent (the turbulence onset phase) \citep{Tambe2021, Tambe2023}. When the inflow Reynolds number $Re_L$ is increased at a fixed $\Omega^*$ (consequently, $S_b$ is decreased), the spiral vortex growth shifts downstream on the cone---showing that the scaled meridional length $l^*/\delta_\nu^*$ does not scale the spiral vortex growth. In contrast, figure \ref{fig:Irms}(b) shows that, on G\"ortler number scale, the temperature fluctuation profiles associated with the spiral vortex growth overlap with each other. This confirms that G\"ortler number is an appropriate parameter for scaling the spiral vortex growth region on a rotating cone in axial inflow.

\begin{figure}
    \centering
    \includegraphics[width=0.65\textwidth]{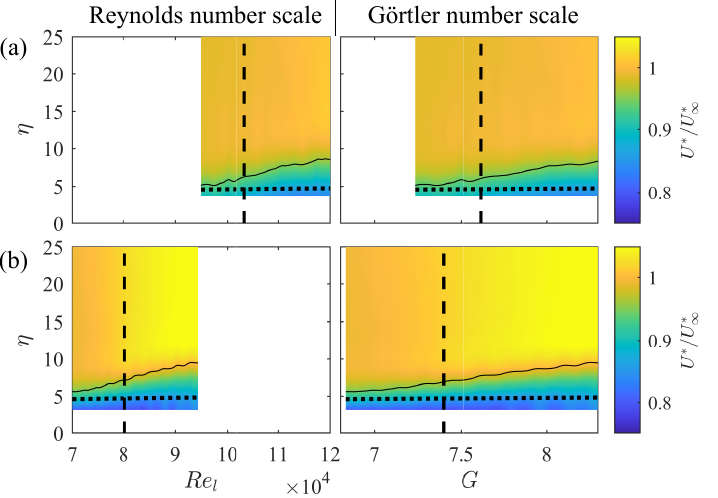}
    \caption{Mean meridional velocity field obtained from PIV over a rotating cone of $\psi=15^\circ$ on Reynolds number and G\"ortler number scales with : (a)~$S_b=1.9$ and $Re_L=9.7 \times10^4$; (b)~$S_b=3.1$ and $Re_L=6.2 \times 10^4$. Solid and dotted black lines represent the boundary layer thicknesses $\delta_{95,\textrm{exp}}$ and $\delta_{95,\textrm{th}}$ obtained from measured mean flow and computed basic flow, respectively. The black dashed line represents the maximum amplification identified from surface temperature fluctuations.}
    \label{fig:15meanU}
\end{figure}

The amplified spiral vortices begin to interact with the outer flow and enhance mixing; the enhanced mixing starts to increase the boundary layer thickness. For example, figure \ref{fig:15meanU}(a) and (b), show the mean meridional velocity fields over a rotating cone ($\psi=15^\circ$) on the local Reynolds number scale, at two different operating conditions. Here, the black dashed line marks the location of maximum amplification identified from the surface temperature fluctuations (e.g. figure \ref{fig:Irms}). The solid and dotted black lines represent the boundary layer thicknesses $\delta_{95,\textrm{exp}}$ and $\delta_{95,\textrm{th}}$ obtained from the measured mean flow and computed basic flow, respectively;  $\delta_{95}$ is the wall-normal extent along $\eta$ up to which the meridional velocity deficit (or excess, e.g. at high $S$ figure \ref{fig:basic flow}(b)) is more than $5\%$ of the outer irrotational flow velocity. Before the maximum amplification, the measured boundary layer thickness (solid line) follows that of the basic flow (dotted line) within $0.1-0.2 \delta_{95,\textrm{exp}}$ but starts to drastically deviate around the maximum amplification. The velocity fields in two different cases, as shown in the left columns of figures \ref{fig:15meanU}(a) and (b), do not align with each other on the local Reynolds number $Re_l$ scale---suggesting that $Re_l$ alone is not an appropriate scaling parameter for the rotating cones. However, on G\"ortler number scale, the near-wall velocity fields and the maximum amplification locations align close to each other at $G\approx7.5$, as shown in the right columns. This further confirms that the G\"ortler number appropriately scales the spiral vortex growth region on a rotating slender cone $\psi=15^\circ$.   

\begin{figure}
    \centering
    \includegraphics[width=0.85\textwidth]{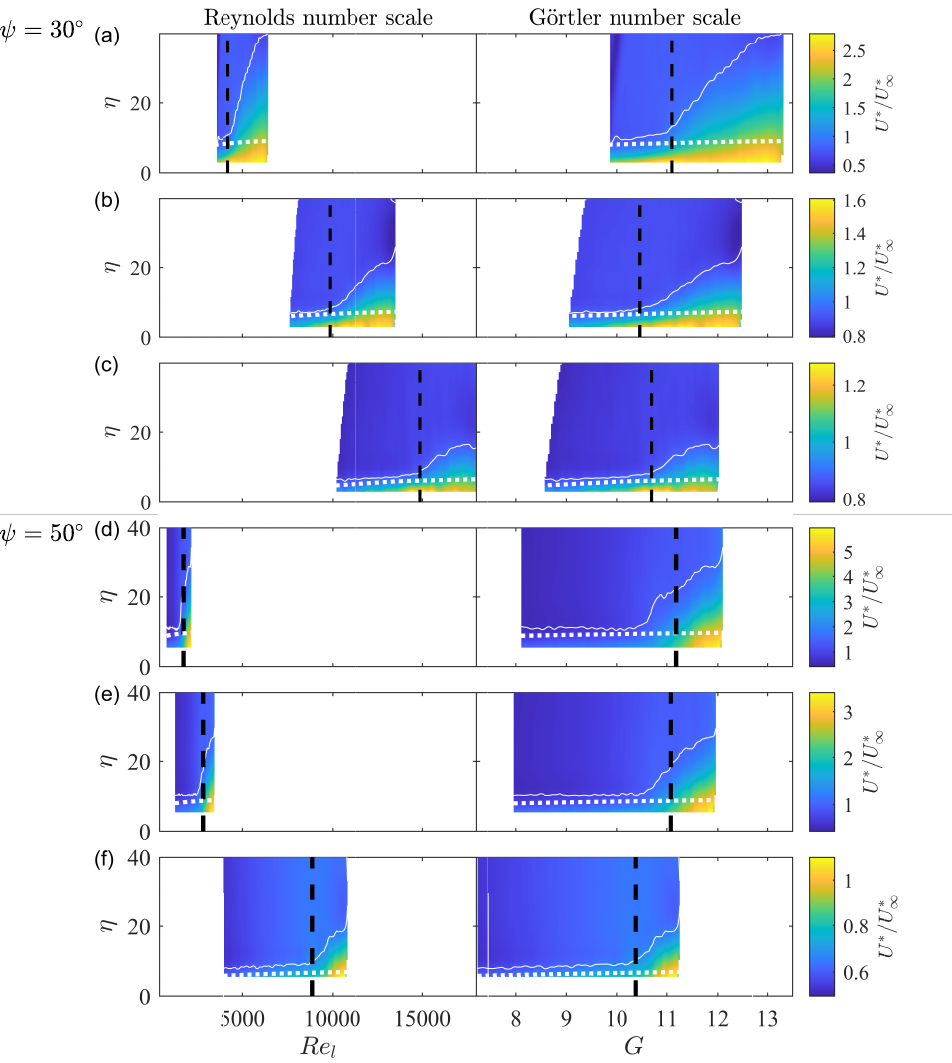}
    \caption{Mean meridional velocity field obtained from PIV over a rotating cone of (a--c) $\psi=30^\circ$ and (d--f) $50^\circ$ on 
    Reynolds number and G\"ortler number with (a) $S_b=28.4$, $Re_L=8 \times 10^3$; (b) $S_b=15.7$, $Re_L=1.5 \times 10^4$; (c) $S_b=10.8$, $Re_L=2.2 \times 10^4$; (d) $S_b=94$, $Re_L=3 \times 10^3$; (e) $S_b=55.6$, $Re_L=5\times 10^3$; and (f) $S_b=18.0$, $Re_L=1.7\times 10^4$. Solid and dotted white lines represent the boundary layer thicknesses $\delta_{95,\textrm{exp}}$ and $\delta_{95,\textrm{th}}$ obtained from measured mean flow and computed basic flow, respectively. The black dashed line represents the maximum amplification identified from surface temperature fluctuations.
    }
    \label{fig:30_50meanU}
\end{figure}

Generally, increasing the half-cone angle has a stabilising effect on the boundary-layer, such that the transition is delayed to higher values of local Reynolds number $Re_l$ and local rotational speed ratio $S$ \citep{Kobayashi1987,Garrett2010, Tambe2023}. For example, at a fixed local Reynolds number ($Re_l$), broader cones require a stronger rotation effect (higher $S$) to cause the boundary layer transition. Figure \ref{fig:30_50meanU} shows the mean velocity fields for two rotating cones ($\psi=30^\circ$ and $50^\circ$, respectively) on $Re_l$ and $G$ scales. Due to the high rotation rates, typically $S \gtrsim 5$, the local meridional velocity is higher as compared to the boundary layer edge (e.g. as also seen in figure \ref{fig:basic flow}(b)). Similar to $\psi=15^\circ$ (figure \ref{fig:15meanU}), on broad cones $\psi=30^\circ$ and $50^\circ$ (figure  \ref{fig:30_50meanU}), the measured boundary layer thickness (solid white line) is close to that of the basic flow (dotted white line) within $0.1-0.2 \delta_{95,\textrm{exp}}$ until the maximum amplification, beyond which it increases. For both these cones, the velocity fields align with each other on the G\"ortler number scale, unlike on the local Reynolds number scale. This shows that the G\"ortler number-based scaling is effective for a range of half cone angles: from slender ($\psi=15^\circ$) to broad ($\psi=50^\circ$) cones. The G\"ortler number related to the maximum amplification increases from around $G\approx7.5$ for a slender cone $\psi=15^\circ$ to around $G\approx10-11$ for the broader cones $\psi=3 0^\circ$ and $50^\circ$.

 \begin{figure}
    \centering
    \includegraphics[width=1\textwidth]{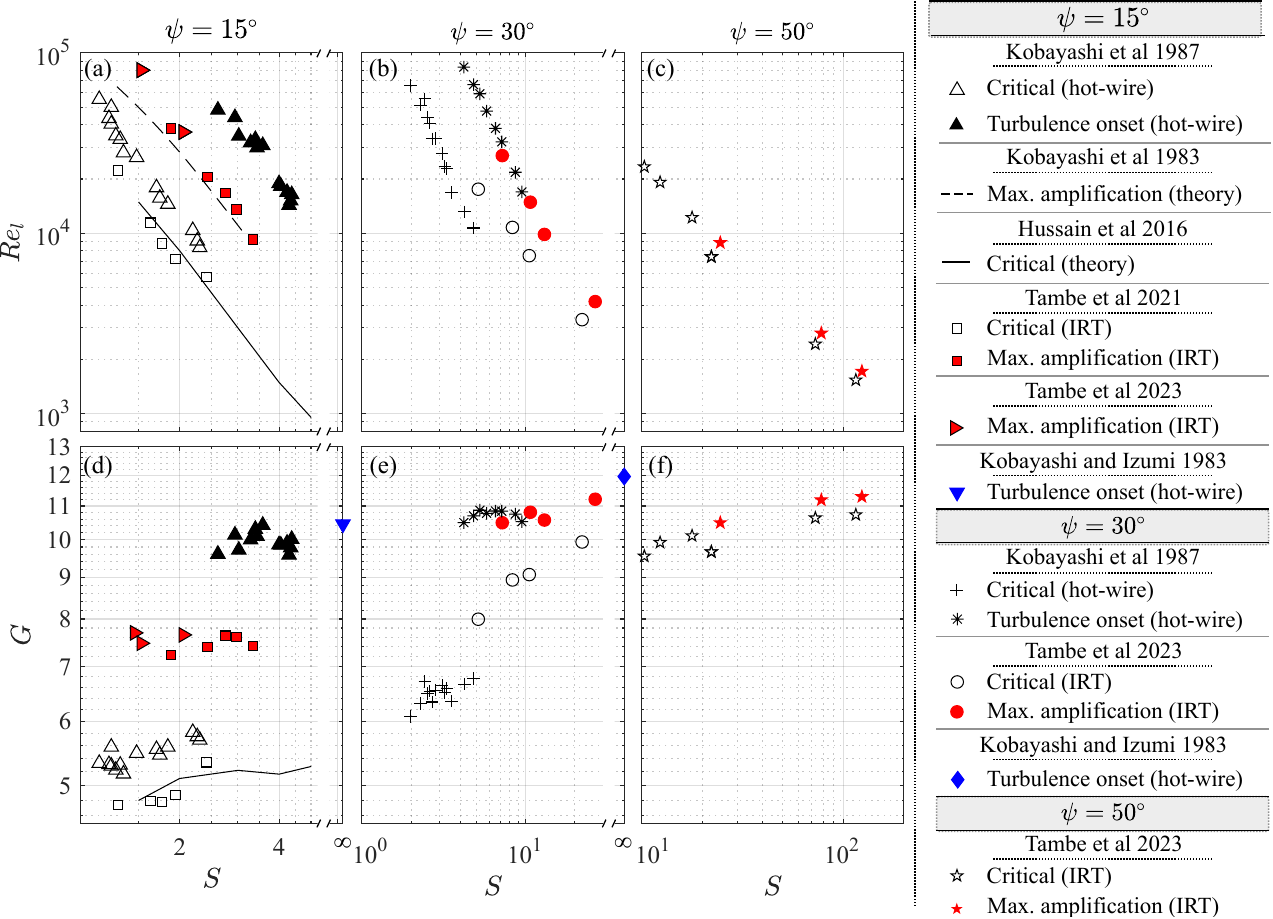}
    \caption{The boundary layer transition on rotating cones ((a,d) $\psi=15^\circ$ , (b,e) $\psi=30^\circ$, and (c,f) $\psi=50^\circ$) in two different parameter spaces: (a, b and c) Reynolds number and local rotational speed ratio ($Re_l-S$) as reported in the literature, and (d, e, and f) the estimated G\"ortler number and local rotational speed ratio ($G-S$). }
    \label{fig:Re_S_G}
\end{figure}

To assess the generality of G\"ortler number-based scaling of rotating-cone boundary-layer transition, G\"ortler numbers are evaluated for the transition points reported by different studies in the literature in $Re_l-S$ parameter space, which used different wind tunnel facilities, different model sizes (base diameters $D^*=L^*\sin \psi=0.047-0.1$m), different measurement techniques (hot-wire anemometry, infrared thermography, etc.), transition criteria, free-stream turbulence levels ($0.05-1\%$), half-cone angles ($\psi=15^\circ$, $30^\circ$, and $50^\circ$) \citep{Kobayashi1983,Kobayashi1983a,Kobayashi1987,Tambe2021,Tambe2023}, and theoretical analysis \citep{Hussain2016}. The transition points (relating to critical, maximum amplification, and turbulence onset phases) are shown in $Re_l-S$ space directly as they are reported in the literature (figures \ref{fig:Re_S_G}(a)-(c)) and, after estimating their respective G\"ortler numbers, they are represented in $G-S$ space (figures \ref{fig:Re_S_G}(d)-(f), transformed using equation (\ref{eq:G}), where $\delta$ is obtained from the similarity solution as shown in figure \ref{fig:basic flow}(c) and $l=\sqrt{Re_l\,S /\sin \psi}$). In $Re_l-S$ space (figures \ref{fig:Re_S_G}(a)-(c)),  all transition points show a nearly log-linear behavior. For a $\psi=15^\circ$ cone (figure \ref{fig:Re_S_G}(a)), a good agreement between theory and different measurements is shown, confirming that $Re_l$ and $S$ together are appropriate to represent the boundary-layer transition region on rotating cones. However, in $G-S$ space (figure \ref{fig:Re_S_G}(d) for $\psi=15^\circ$), the transition points appear at respectively fixed G\"ortler numbers---regardless of the local rotational speed ratio in the investigated range $S\gtrsim1$. The G\"ortler numbers for the measured turbulence onset points by \citet{Kobayashi1987} in axial inflow $S\approx2.5-4.5$ agree with that of \citet{Kobayashi1983a} in still fluid $S\approx \infty$---both studies used the same measurement technique. Moreover, the critical points predicted by \citet{Hussain2016} also appear within a narrow range of G\"ortler number. Considering the small variation of the respective G\"ortler numbers relative to the uncertainty around $\pm0.1G$, we can conclude that the critical and the maximum amplification points as well as turbulence onset on the rotating $15^\circ$ cone are scaled by G\"ortler number regardless of the axial inflow for $S \gtrsim 1$. At low local rotational speed ratio, i.e. $S<<1$, the centrifugal effects are expected to be weak and different mechanisms dominate transition \citep{Song_Dong_Zhao_2023}. However, at $S \gtrsim 1$, the centrifugal instability is known to be dominant on the slender cone~\citep{Kobayashi1983,Kobayashi1983a,Hussain2016,Kato2021}, and the influence of $S$ on G\"ortler number is reported here for the first time. Thus, when the transition is induced due to the strong rotation effect ($S \gtrsim 1$),  G\"ortler number is an appropriate parameter to scale the boundary-layer transition on a rotating slender cone $\psi=15^\circ$ rather than using the two-parameter space $Re_l-S$.

For broader cones $\psi=30^\circ$ and $50^\circ$, figures \ref{fig:Re_S_G}(e) and (f) show the transition points in $G-S$ space, respectively. Unlike their near-log linear trends in the $Re_l-S$ space (figures \ref{fig:Re_S_G}(b) and (c)), the maximum amplification and turbulence onset points in $G-S$ space appear in a narrow G\"ortler number range for the investigated values of $S$. Moreover,  maximum amplification points for both the cones ($\psi=30^\circ$ and $50^\circ$)  appear in the range $G\approx 10-11$ as shown in figures \ref{fig:30_50meanU} and \ref{fig:Re_S_G}(e) and (f). This range agrees with the results of \citet{Kato2021} on a $\psi=30^\circ$ cone rotating in still fluid ($S\approx\infty$), although there are some differences in the transition criteria and the way of calculating the momentum thickness; \citet{Kato2021} reported a gradual thickening of the boundary-layer, starting around $G=10$ based on the measured momentum thickness evaluated below the $90\%$ boundary-layer thickness whereas the present $\delta$ is obtained by integrating the similarity solution in the infinite space. Moreover, the turbulence onset points for $\psi=30^\circ$ cone appear at $G=10-12$ (figure \ref{fig:Re_S_G}(e)). Thus, the maximum amplification point, beyond which the measured mean flow drastically deviates from the similarity solution flow, and turbulent onset occur in a well-defined range of G\"ortler numbers for a wide range of investigated $S$. However, the critical points vary with respect to the local rotational speed ratio $S$ in $G-S$ space at high rotational speed ratio $S\gtrsim 5$ (figure \ref{fig:Re_S_G}(e) and (f)),  where, for broader cones $\psi \geq 30^\circ$, the primary crossflow instability dominates the flow rather than centrifugal instability \citep{Kobayashi1983a,Kobayashi1987,Garrett2010}. 

It is interesting that G\"ortler number dominates the maximum amplification and turbulence onset on broader cones $\psi=30^\circ$ and $50^\circ$ at higher $S \gtrsim 5$ (figures \ref{fig:30_50meanU} and \ref{fig:Re_S_G}(e) and (f)), where the primary instability is the crossflow instability and forms co-rotating primary vortices. A possible explanation for this might be that the centrifugal effects cause a secondary instability; some measurements, e.g., figure 7 from \citet{Tambe2023} and also the top right of figure 8b from \cite{Kobayashi1983a}, show that new counter-rotating vortices emerge near the maximum amplification (after the primary co-rotating vortices have developed), which might be caused by centrifugal effects. Dominance of G\"ortler number in this region suggests that the centrifugal effects play an important role in the spiral vortex amplification and turbulence onset even on the broad cones where the crossflow primary instability dominates the flow initially.

\section{Conclusion}

The G\"ortler number is found to scale the centrifugal instability-led boundary-layer transition on a rotating slender cone ($\psi=15^\circ$) in axial inflow. For the local rotational speed ratio $S \gtrsim 1$, the critical, maximum amplification, and turbulence onset points appear at well-defined G\"ortler numbers respectively, regardless of the axial inflow strength or local rotational speed ratio $S$. Therefore, the G\"ortler number alleviates the need to use the conventional two-parameter space of local Reynolds number $Re_l$ and local rotational speed ratio $S$ to represent the transition points on a rotating slender cone ($\psi=15^\circ$) in axial inflow for $S \gtrsim 1$.

On broader cones $\psi=30^\circ$ and $50^\circ$, where the crossflow instability is the dominant primary instability, the maximum amplification and turbulence onset are found to occur around $G=10-12$ which is affected marginally by $S$, although the critical G\"ortler number varies with $S$. This suggests that, for $S \gtrsim 1$, the centrifugal effects play an important role in boundary-layer transition for a wide range of investigated rotating cones with $15^{\circ} \leq \psi \leq 50^{\circ}$, regardless of the axial inflow. Further investigation is required to understand the detailed role of the centrifugal effects in the turbulence onset mechanism.

\backsection[Supplementary data]{\label{SupMat}Supplementary material and movies are available at \\https://doi.org/10.1017/jfm.2019...}

\backsection[Acknowledgements]{
Authors wish to acknowledge Ferry Schrijer, Arvind Gangoli Rao, Leo Veldhuis, and TU Delft Wind tunnel labs for the experimental data. S. Tambe acknowledges the fellowship support of Department of Science and Technology, Government of India for this work and European Union Horizon 2020 program: Clean Sky 2 Large Passenger Aircraft (CS2-LPA-GAM-2018-2019-01), and CENTERLINE (Grant Agreement No. 723242) for the experimental data. K. Kato acknowledges support from JSPS KAKENHI Grant Number JP22K20406. 
}

\backsection[Declaration of interests]{{\bf Declaration of Interests}. The authors report no conflict of interest.}

\backsection[Data availability statement]{Data is available upon a reasonable request.}

\backsection[Author ORCIDs]{Sumit Tambe:  0000-0002-2628-4051, Kentaro Kato: 0000-0002-5532-2379, Zahir Hussain: 0000-0001-6756-6058}

\backsection[Author contributions]{S. Tambe contributed to measuring and analysing data, conceptualising, and writing the first draft manuscript. K. Kato contributed to conceptualising and writing the manuscript. Z. Hussain contributed to basic flow computation, conceptualising, and writing the manuscript.}




\bibliographystyle{jfm}
\bibliography{Refs}

\end{document}